\documentclass{article}

\usepackage{microtype}
\usepackage{graphicx}
\usepackage{subfigure}
\usepackage{booktabs}
\usepackage{amsmath} 
\usepackage{mathtools}
\usepackage{amssymb}
\usepackage{enumitem}

\usepackage{hyperref}

\usepackage[accepted]{icml2019}

\begin{document}

\twocolumn[

\title{Fine-grained large-scale content recommendations for MSX sellers}
\date{\vspace{-0.2in}}
\maketitle

\icmlsetsymbol{equal}{*}

\begin{icmlauthorlist}
\icmlauthor{Manpreet Singh}{MS}
\icmlauthor{Ravdeep Pasricha}{MS}
\icmlauthor{Ravi Prasad Kondapalli}{MS}
\icmlauthor{Kiran R}{MS}
\icmlauthor{Nitish Singh}{MS}
\icmlauthor{Akshita Agarwalla}{MS}
\icmlauthor{Manoj R}{MS}
\icmlauthor{Manish Prabhakar}{MS}
\icmlauthor{Laurent Bou\'e}{MS}
\end{icmlauthorlist}

\icmlaffiliation{MS}{Microsoft, CX Data Cloud + AI}

\vspace{0.4in}

\begin{abstract}

\vspace{0.4in}

One of the most critical tasks of Microsoft sellers is to meticulously track and nurture potential business opportunities through proactive engagement and tailored solutions.  Recommender systems play a central role to help sellers achieve their goals.  In this paper, we present a content recommendation model which surfaces various types of content (technical documentation, comparison with competitor products, customer success stories etc.)  that sellers can share with their customers or use for their own self-learning.  The model operates at the opportunity level which is the lowest possible granularity and the most relevant one for sellers.  It is based on semantic matching between metadata from the contents and carefully selected attributes of the opportunities.  Considering the volume of seller-managed opportunities in organizations such as Microsoft, we show how to perform efficient semantic matching over a very large number of opportunity-content combinations.  The main challenge is to ensure that the top-5 relevant contents for each opportunity are recommended out of a total of~$\approx 40,000$ published contents.  We achieve this target through an extensive comparison of different model architectures and feature selection. Finally, we further examine the quality of the recommendations in a quantitative manner using a combination of human domain experts as well as by using the recently proposed ``LLM as a judge'' framework.

\vspace{0.4cm}

\end{abstract}
\vspace{0.4in}
]

\icmlcorrespondingauthor{\\ Laurent Bou\'e}{laboue@microsoft.com}
\printAffiliationsAndNotice{}

\section{Introduction}
\label{sec:introduction}

In large software organizations, sellers have to nurture, cultivate and maintain relationships with a large ecosystem of partners, customers and dependent stakeholders.  At Microsoft, sellers use the Microsoft Sellers Experience - MSX tool to navigate this intricate landscape.  Recently, more and more Copilot-inspired systems have been integrated into~MSX to help guide sellers and improve their productivity.  However, there had not yet been solutions that operate at the lowest level of granularity that sellers work at on a daily basis: the opportunity level. 

In a CRM system, an ``opportunity'' refers to a potential revenue-generating event or transaction that arises during the course of managing customer relationships. It represents a chance for a business to convert a lead into a customer, close a deal, or expand its services to an existing customer. Opportunities are pivotal moments in the sales process that require careful nurturing and management to maximize the likelihood of success.  Sharing the right content with the customer at each sales stage is one of the key factors that helps in moving the opportunity to the next stage. It is important to share only a few but relevant documents (pitch decks, customer success stories, battle cards to show the comparison with competitors...) with the customer to help them quickly understand the value of Microsoft products.

In this paper, we show how we have built an opportunity-level recommender system whose purpose is to present the sellers with the top-5 technical documents drawn from the Seismic content repository~\cite{seismic} to increase the sale velocity.  Comprising of a very large catalog of technical documentation, product descriptions, customer success stories and more, Seismic is a leading content management system widely used across businesses to manage their digital content.  In our case, we consider a catalog of approximately~$\approx 40,000$ unique documents.  Since those documents are decided upon within the context of a specific opportunity, the goal is that they can be shared by the sellers to their customers in order to move the opportunity towards closure in a more targeted manner.  Other than a simple rule-based engine which surfaces too many contents, sellers do not currently have any automated guidance as to which Seismic documents would be good candidates to share with their clients.

We start in Section~\ref{sec:semanticMatching} by discussing how we have formulated the recommender system as a form of semantic matching between opportunities and Seismic documents.  Due to the large volume of opportunities, we discuss how we have designed our solution for large-scale semantic matching.  Next, we discuss in Section~\ref{sec:perf} how we evaluate the quality of the recommendations using a mixture of different techniques to alleviate the fact that there is no ground-truth data.  Finally, we illustrate in~Section~\ref{sec:MSXintegration} how our solution has been integrated into MSX and is currently being used by real-world Microsoft sellers.

\section{Large scale semantic matching for content recommendations}
\label{sec:semanticMatching}

As mentioned in the introduction, the objective of this work is to recommend relevant Seismic documents given the context of a specific~MSX~opportunity.  

With the democratization of large language models, semantic search has become a well-established technique and we describe in this section how the recommender model can be formulated in these terms.  Generally, this formulation differs from traditional recommender systems~\cite{traditionalReco} that rely on user-item interactions to make predictions.  Collaborative filtering and nearest-neighbor methods analyze users’ behavior and preferences to find similarities and identify recommendations.  In contrast, our formulation as a semantic matching method relies on natural language understanding of the description of the Seismic documents and of the sellers.

\subsection{Prompt engineering}
\label{sec:prompt}

To achieve this we follow the approach initiated in~\cite{copilotMSX} where it was shown that it is possible to produce high-quality recommendations between Seismic documents and the end user of the recommendations.  In this previous study, the end user was a real time chatbot conversation agent whereas here it is the context associated with specific opportunity.  We use the same technique for metadata prompt engineering based on metadata of the Seismic documents and metadata of the opportunities.  Essentially, the idea boils down to summarizing the Seismic documents and the opportunities into a short text-based description that contains the most important attributes.

In case of Seismic documents we use features like ``name'', ``description'', ``solution area'', ``product'' etc. Similarly for opportunities we look those features which may match those of the Seismic documents as much as possible.  By  using features which are common on both sides of the semantic matching, we maximize the chance of successful recommendations.  The idea is to capture the most common features on both sides so that the language model based embeddings provide relevant documents for opportunities.   More details about the architecture can be found in Section~\ref{sec:architecture}.

\subsection{Model architecture}
\label{sec:architecture}

\begin{figure*}[ht]
\begin{center}
\centerline{\includegraphics[width=1.7\columnwidth]{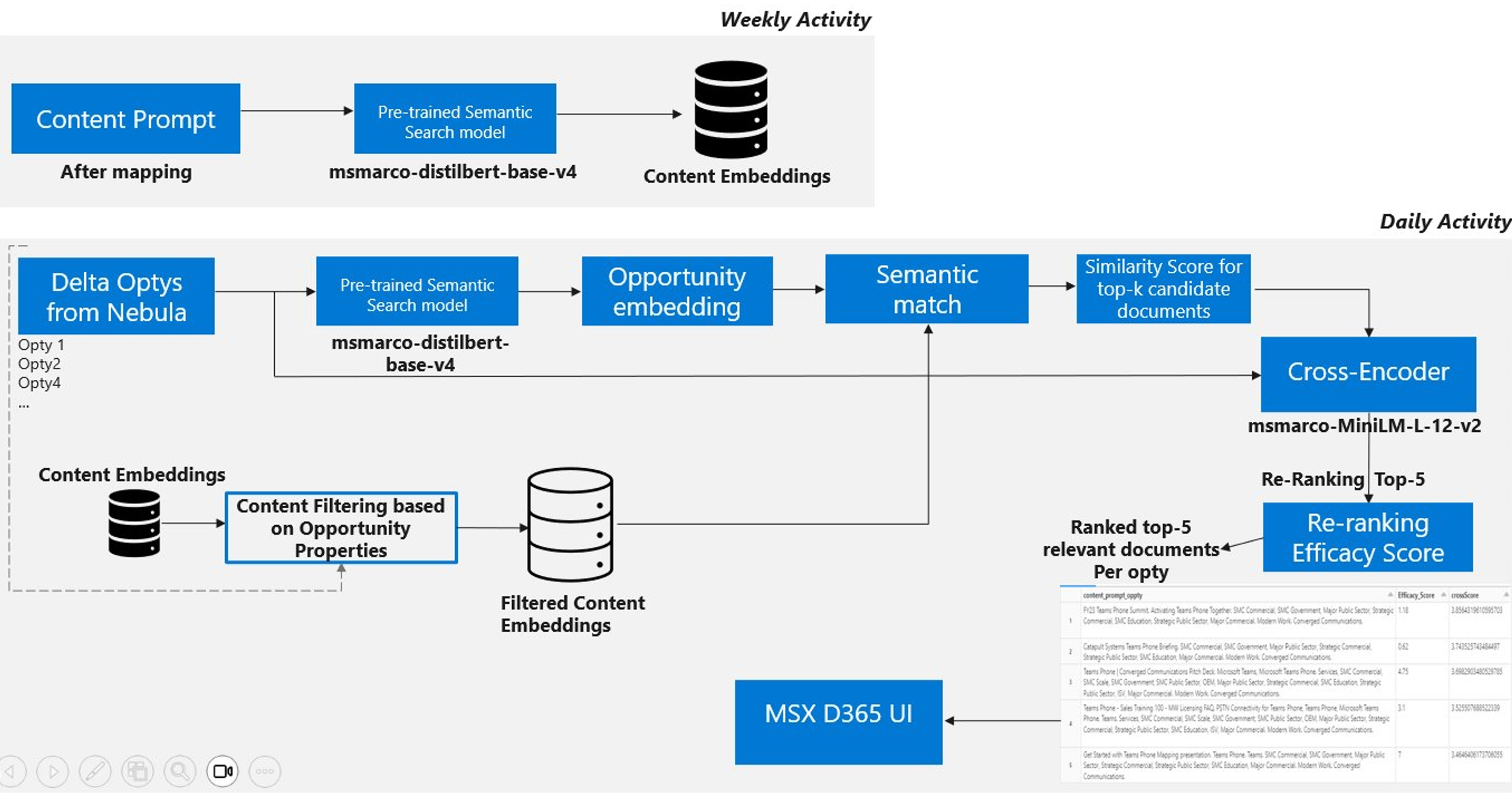}}
\caption{{\bf Top)} Seismic documents are summarized into textual descriptions, referred to as ``content prompts'' based on their metadata.  These prompts are then run through a \texttt{DistillBERT} language model~\cite{sanh2019distilbert} (pre-trained on \texttt{MSMarco} dataset).  Those embeddings are refreshed on a weekly basis.  {\bf Bottom)} The $\delta$-opportunities (defined in the main part of the text) are gathered from Nebula~\cite{nebula}, which is an in-house~ETL system developed by the~SPS team.  Next, the opportunity prompt (summarized attributes of the opportunity into a textual prompt in a manner similar to content prompts) is run through the same \texttt{DistillBERT} language model and compared to content embeddings to generate a list of top-50 candidate documents.  Those candidates are re-ranked using the~\texttt{MSMarco MiniLM} pre-trained cross-encoder.  Finally, the top-5 results for each opportunity are stored in ADLS from where the Nebula insights pipeline pushes the results to a Cosmos database. The~.NET API then pulls the recommended Seismic documents per opportunity from Cosmos and displays it in the UI (see Section~\ref{sec:MSXintegration}).}
\label{fig:modelArchitecture}
\end{center}
\end{figure*}

The architecture of the model itself is inspired by the one designed for the real-time Copilot recommender system previously developed and already deployed in MSX production~\cite{copilotMSX}.  

The idea consists of leveraging a 2-stage system for fast (but slightly inaccurate) retrieval of the top-50 relevant documents (for each opportunity) followed by a much slower step of re-ranking using a cross-encoder model.  Generally, both language models are pre-trained on the \texttt{MS MARCO} dataset~\cite{bajaj2016ms} which is known to produce good embeddings for these types of question-answering systems.  For more technical details regarding the data flow, we refer the reader to~Fig.~\ref{fig:modelArchitecture} and to~\cite{copilotMSX} for the choice of parameters.  One important difference from the Copilot model of~\cite{copilotMSX} is that the current model is delivered in a daily-refreshed batch mode instead of real-time.  

Because of the volume of opportunities for which the recommender system is making predictions, we have decomposed the operations of the model into three distinct parts:
\begin{itemize}
\item A one-time pre-population of the recommendations for the entire dataset of~$\approx 700,000$ opportunities representing the last~6 quarters of open opportunities.
\item A daily batch-mode refresh of the~$\delta$-opportunities.  Opportunities are classified as~$\delta$-opportunities if they are net-new opportunities or if some critical properties have changed in the last~24 hours since the last refresh~\footnote{In our case, those properties are opportunityId, opportunityname, salesplay, salesstagename, primaryproduct, segment, areaname.}. In practice, we deal with~$\approx 10,000$ such opportunities on a daily basis.
\item Furthermore, in order to keep up with changes in the Seismic catalog, we refresh the content embeddings on a weekly cadence.  The embeddings are stored in~ADLS but brought up to memory in an Azure Databricks Spark cluster for the daily opportunity refresh.  This activity is a separate module and its frequency can be increased as required.
\end{itemize}

\subsection{Orders of magnitude}

Let us now turn our attention to the another aspect related to the scale of the semantic search at play here.

On one hand, the total number of opportunities with   ``open status'' in~FY23 and~FY24 (i.e. last rolling 6 quarters) is~$\approx 700,000$.  On the other hand, the total number of published documents is~$\approx 40,000$.  Na\"ively, this means that one should consider all~$\approx 700,000 \times 40,000$ opportunity-content pairs.  This would result in~$ \approx 28 \times 10^9$ combinations.   Clearly, one must introduce some filters in order to reduce this large number of combinations.

In the following, we use three features (``sales stage'', ``area'' and ``solution area'') as filters.  This allows us to reduce the content search space to roughly~$\approx 7,000$ Seismic documents.  Now we have~$\approx 5 \times 10^9$ combinations reducing the overall computational effort required by~$\approx 80\%$.  Nonetheless, this still leaves us with multiple billions of combinations to perform so this still remains a very large scale semantic matching problem.

\subsection{Run-time performance optimization}
\label{sec:runTime}

\begin{figure}[ht]
\begin{center}
\centerline{\includegraphics[width=\columnwidth]{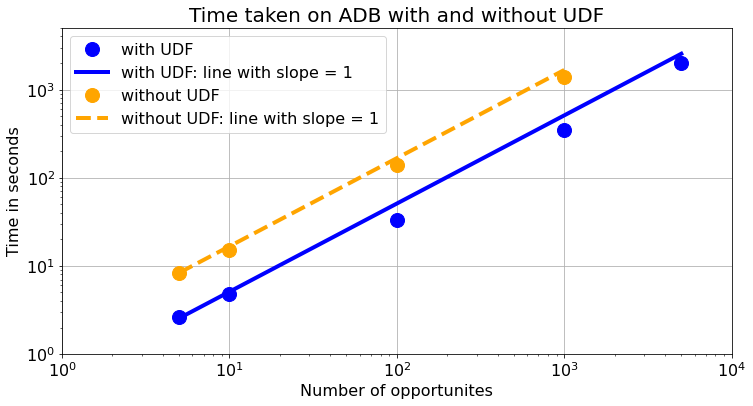}}
\caption{Illustration of the performance gain by using Pandas UDFs on Azure Databricks Spark clusters. As expected, the processing time grows linearly with the number of opportunities.  Further incremental gains may be obtained by increasing the size of the Spark cluster.  Note that the number of opportunities is not the same as the number of records processed by the cross-encoder. Consider, for instance, that we have $1,000$ opportunities.  In that case, the total number of records processed by cross encoder is $50 \times 1,000 = 50,000$ where the factor of~50 corresponds to the number of candidates retrieved in the first stage before re-ranking.}
\label{fig:runtimePerf}
\end{center}
\end{figure}

In order to reduce the computational complexity, an assumption is to consider opportunities as all independent from each other.  (We discuss the limitations associated with this assumption in the conclusion.)  Under this assumption, generating recommendations for all the opportunities becomes a fundamentally ``embarassingly-parallel'' task.

Simple profiling reveals that it is the cross-encoder re-ranking stage which is overwhelmingly the most time-consuming part of the architecture described in~Fig.~\ref{fig:modelArchitecture}.  The total number of records that needs to go through this re-ranking is~$ 50 \times$ the number of opportunites since we retrieve 50 candidates for each opportunity.  Each record is a pair of prompts: a content prompt shortlisted by the bi-encoder retriever and an original opportunity prompt.

Pandas User Defined Functions - UDFs \cite{pandasUDF} in PySpark are a popular technique to combine complex data transformation pipelines leveraging Python libraries that may not be natively available in Spark with the convenient parallelism offered by workloads on Spark cluster infrastructures. Accordingly, we utilize Pandas UDFs to distribute the computation across on an Azure Databricks  cluster.  Figure \ref{fig:runtimePerf} shows the time taken to process number of opportunities with and without Pandas UDFs.  This confirms that Pandas UDFs on Spark clusters lead to a consistent gain in performance. 

Note that incremental further gains may be obtained by proportionately increasing the size of the allocated Spark clusters. Using a Spark cluster with 96 vcores as required for production deployment of the model, we have reduced the processing time from~$\approx 2$s per opportunity to~$\approx 90$ms.

\section{Relevance/performance evaluation of the recommendations}
\label{sec:perf}

Evaluating the quality of recommender systems without ground-truth data poses a challenge due to the absence of a specific objective criteria for assessment.  Consequently, it is common to rely on human experts to provide subjective evaluations.  Even though this process does offer valuable insights, it can be prohibitively labor-intensive and costly. 

We start in Section~\ref{sec:crossProxy} by preparing a set of evaluation queries completed by~3 human domain experts and show how the cross-encoder scores produced by the model are in good agreement with the scores given by those experts.  Next, we carry out in Section~\ref{sec:ablation} an ablation study to investigate the relative importance of the features used in the prompts.  Finally, we explore the LLM as a judge framework where GPT-4 is used as an independent evaluator in Section~\ref{sec:LLMjudge}.

\subsection{Human expert evaluation and cross-encoder scores as a proxy}
\label{sec:crossProxy}

\begin{figure}[ht]
\begin{center}
\centerline{\includegraphics[width=\columnwidth]{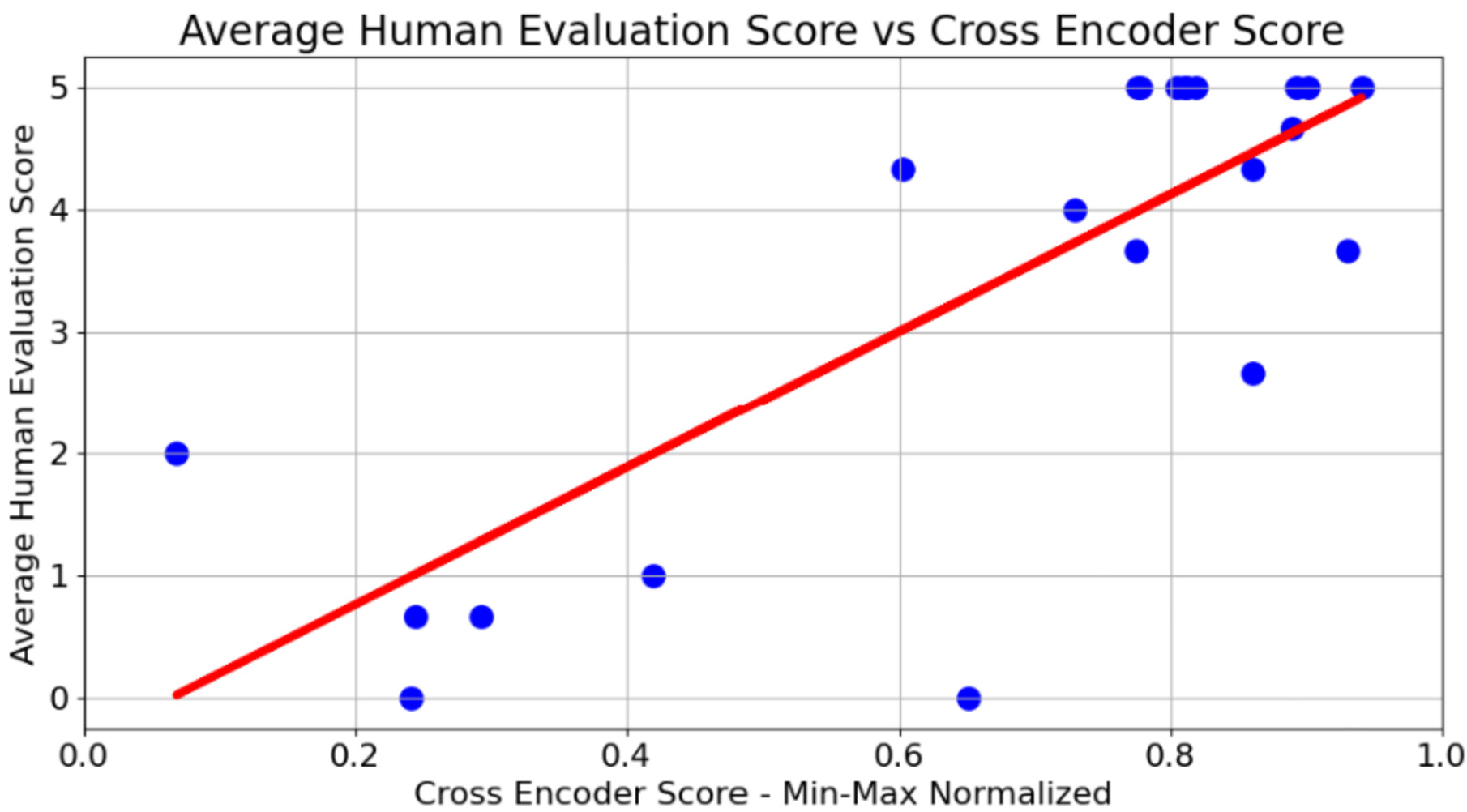}}
\caption{Illustration of the good alignment between cross-encoder scores and human judgment with a Pearson correlation coefficient of~0.78.  Going beyond linear correlations, we have also estimated a rank-based Spearman's coefficient of~0.64.}
\label{fig:crossEncoder_proxy}
\end{center}
\end{figure}

We curated a list of 22 queries to be sent for evaluation by~3 human experts.  For each of these queries, the recommended Seismic documents are predicted by the model and the top-5 results are presented to the evaluators who are asked to give them a rating from~0 to~5 (higher is better).  Once those scores are provided by the experts, we ask the question of how much does the cross-encoder score align with human judgments.

As presented in Fig.~\ref{fig:crossEncoder_proxy}, we see that there is indeed a strong positive Pearson correlation coefficient of~$0.78$ between the cross-encoder score and the scores produced by human experts (averaged over the 5 recommended items).  This agreement is very good to separate bad recommendations from the better ones.  This means that the cross-encoder score can realistically be used, at least in a binary classifier manner to quickly identify good vs. bad recommendations.  Those can then be processed along towards human experts for further analysis with an improved ``triage'' time.

In the absence of ground-truth data, those correlation results based on Pearson and Spearman's statistics are encouraging.  Furthermore, we note that the human reviewers confirmed that the documents they expected to see as a response to their queries consistently showed up in the top-5 recommendations.  This gives an indication that the recall rate (although not directly measurable without ground-truth data) is at least acceptable from the stakeholder's perspective.

\subsection{Ablation study}
\label{sec:ablation}

\begin{figure}[ht]
\begin{center}
\centerline{\includegraphics[width=\columnwidth]{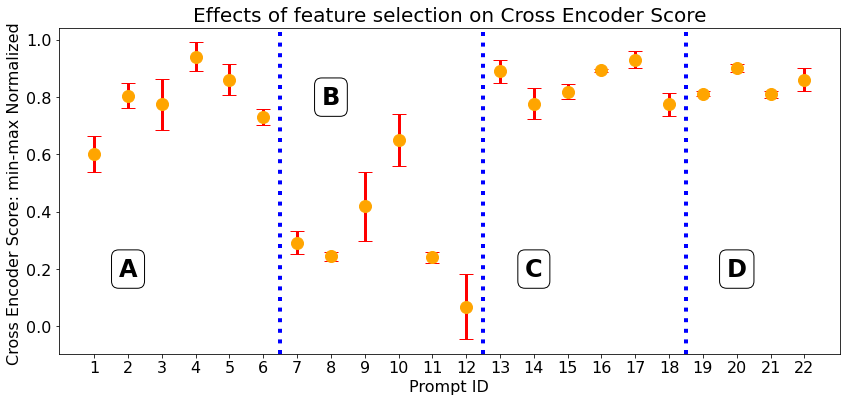}}
\caption{Cross-encoder scores (as returned by the model) vs. each one of the~22 evaluation prompts.  The dashed vertical lines represent the different groups~{\bf A},~{\bf B},~{\bf C}, and~{\bf D}.  As explained in the main part of the text, the low performance of group~{\bf B} indicates the importance of \small{\textsf{``sales play''}} as a critical feature.}
\label{fig:promptFeatures}
\end{center}
\end{figure}

Now that we have established cross-encoder scores as a good proxy for human judgment of the relevance of the recommendations, we can use these model-produced scores, to assess the relative importance of the features used in the prompts (see Section~\ref{sec:prompt}).  For the purpose of the analysis, we have divided the evaluation queries into~4 groups depending on the features that were used to design the prompts.  \underline{Crucially}, note that the human evaluators were never aware of the underlying division of the queries into~4 groups so this division did not influence their scores (nor were the evaluators aware of the cross-encoder scores). 

We denote by~$\mathcal{F}_\cap$ the set of~3 features that were used both on the documents as well as on the opportunity prompts:
\begin{equation*}
\mathcal{F}_\cap = \big\{ \small{ \textsf{``sales play'', ``solution area'', ``product''} } \big\}
\end{equation*}
The~22 evaluation queries are divided into~4 groups consisting of:
\begin{enumerate}[label=\textbf{\Alph*}]
\item All members of~$\mathcal{F}_\cap$ are used in the prompts.
\item Remove 1 feature so that features from \\ $\mathcal{F}_\cap \setminus \small{\textsf{``sales play''}}$ \\ are used in the prompts.
\item Remove 1 feature so that features from \\ $\mathcal{F}_\cap \setminus \small{\textsf{``product''}}$ \\ are used in the prompts.
\item Remove 2 features so that features from \\ $\mathcal{F}_\cap \setminus \ \small{\textsf{``solution area''}} \setminus \small{\textsf{``product''}} $ \\ are used in the prompts.
\end{enumerate}

As can be seen in~Fig.\ref{fig:promptFeatures}, it is clear that \textsf{{\small ``sales play''}} is a critical feature without which the performance of the recommendations is severely affected.  Indeed although groups~{\bf A},~{\bf C} and~{\bf D} have similar performance, group~{\bf B} (i.e. the only one for which the \textsf{\small{``sales play''}} is missing) has a significantly lower cross-encoder for almost all of its queries.  This confirms the importance of \textsf{{\small ``sales play''}} which, as a short descriptive text, (such as ``accelerate innovation with low code'' or ``optimize finance and supply chain'' for example) provides a good opportunity to match Seismic documents and opportunities more precisely.  This also indicates that improving upon this feature and advising the sellers to make \textsf{{\small ``sales play''}} even more descriptive will have a beneficial impact on the relevance of the recommendations.

\subsection{LLM as a judge}
\label{sec:LLMjudge}

\begin{figure}[ht]
\begin{center}
\centerline{\includegraphics[width=\columnwidth]{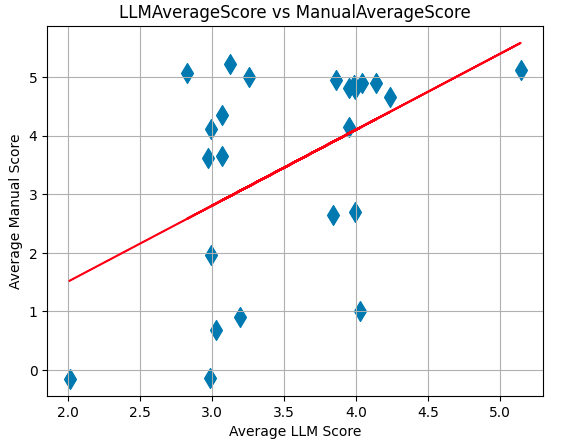}}
\caption{Correlation between average human experts scores and scores as judged by GPT4 for the same~22 evaluation queries.  The red line corresponds to a linear fit with a Pearson correlation coefficient of~0.42 confirming the positive correlation between the different sets of scores. We have also estimated a Spearman's coefficient of~0.57.}
\label{fig:LLMjudge}
\end{center}
\end{figure}

A growing trend in the literature has been the rise of leveraging LLMs as tools to evaluate the performance of tasks solved by other LLMs.  Early results have indicated that strong LLM-based evaluators give scores that are in general alignment with those provided by human domain experts~\cite{LLMjudge, AlpacaFarm, llmEmpirical}.

Following this line of research, we have devised a simple prompt template so that GPT-4 is asked to give its own score for the same set of~22 curated evaluation queries.  The actual prompt template is repeated below: \\
\hrule \vspace{0.1cm} \hrule \vspace{0.3cm}
\texttt{"role" : "You are an AI assistant that helps people find information". \\ \\ "role" : "user" ; Given the following query about an opportunity: 
\begin{itemize}
    \item \textbf{Opportunity Prompt}
\end{itemize}
And the following documents:
\begin{itemize}
    \item \textbf{Doc[1]}, \textbf{Doc[2]}, \textbf{Doc[3]}, \textbf{Doc[4]}, \textbf{Doc[5]}
\end{itemize}
Please perform the following tasks:
\begin{itemize}
\item Calculate the similarity score between the query and each document. The similarity score should reflect how relevant each document is to the information contained in the query. Use a scale from 0 to 5, where 5 indicates a perfect match and 0 indicates no relevance.
\item Provide a brief justification for the ranking based on the similarity scores. \\
\end{itemize}}
\hrule \vspace{0.1cm} \hrule \vspace{0.3cm}
\noindent where \textbf{Opportunity Prompt} is replaced by the actual opportunity prompt and the~\textbf{Doc[i]} are replaced with the recommended content for this opportunity as returned by the model.

As can be seen in~Fig~\ref{fig:LLMjudge}, we do indeed observe a weak correlation between the human expert scores and the scores returned by GPT-4 with a Pearson correlation coefficient of~0.42.

One challenge we have noticed is that GPT-4 seems ``hesitant'' to give very low scores even if its textual justification for the scores indicates a poor match between query and documents (this can be detected by seeing words such as ``however'', ``despite''...).  This behavior has been observed anecdotally by others though we have not been able to find a dedicated study to this form of ``politeness'' bias.

We have also tried different versions of the prompt by putting more emphasis on different aspects (such as rewarding more correct answers or enforcing strict constraints...).  Eventually, we converged a prompt design that generated fair scores and refined it even more using~GPT-4 to converge to the prompt shown above.

\section{Integration in MSX}
\label{sec:MSXintegration}

This new recommender model is fully integrated with a pre-existing Copilot interface previously developed in~\cite{copilotMSX}.  The end result, as it is seen by Microsoft sellers, is illustrated  in~Fig.~\ref{fig:MSXUI} in the appendix.

Whenever a seller opens an opportunity, he/she gets the option to see the recommended contents which can be shared (customer ready) or used by sellers for their private knowledge (i.e. confidential documents that cannot be shared with customers but may still be of value to the sellers). Customer ready contents can be sent to the customer from this window directly. The seller's email will be composed and ``livesendLink'' links for the selected contents will be generated automatically.  Sellers also get the option to see their own past history with the Seismic documents for an opportunity.  They are also offered the possibility to provide feedback about the quality of the recommendations.  The ``search from all contents'' feature allows the sellers to search from all the content space in case they want to search for different/more contents than the recommended ones. This search functionality is powered by the real-time Copilot model~\cite{copilotMSX}.

\section{Conclusion}

The recommender model described in this paper is now undergoing its pilot phase where the purpose is to gather feedback from real-world sellers before it is released to the entire MSX community.  This phase is expected to last for the next~2 quarters during which we plan to gather telemetry about the documents shared by the sellers.  

Meanwhile, there are a number of points we are planning to investigate for further model improvements.  For example, it would be interesting to fine-tune the sentence transformer models using different loss functions such as multiple negative ranking loss.  In the future, we also plan to incorporate feedback from the sellers to improve the quality of the recommendations.  Additionally, programmatic access to the actual content of the Seismic documents is still not available.  It is our intention to take this data source into consideration as soon as it is available.  This will require to modify the prompt engineering strategy discussed in Section~\ref{sec:prompt} into a more elaborate multi-modal system.  Finally, we discussed in~Section~\ref{sec:runTime} how we have treated each opportunity as independent of each other.  Although this simplification was critical for computational efficiency reasons, it potentially limits the quality of the recommendations.  For example, when different opportunities are managed by the same seller, it would be interesting to incorporate seller-specific features and dependencies between the related opportunities.  We plan to study this personalization aspect in a future iteration of the model.

\section{Acknowledgements}

We thank our colleagues in CX Data for their feedback and support. In particular, we thank the CX Data ML Review board (Vijay Agneeswaran, Ivan Barrientos, Jane Huang and Hunny Mehrotra) for reviewing the model and giving us useful inputs. We thank Binh ``Binnie'' Tran from the SPS product management team who helped us with domain knowledge on the Seismic platform and the content space.

\bibliographystyle{ieeetr}
\bibliography{MainText}

\appendix
\section{Appendix on MSX integration}

\begin{figure*}[ht]
\begin{center}
\centerline{\includegraphics[width=2\columnwidth]{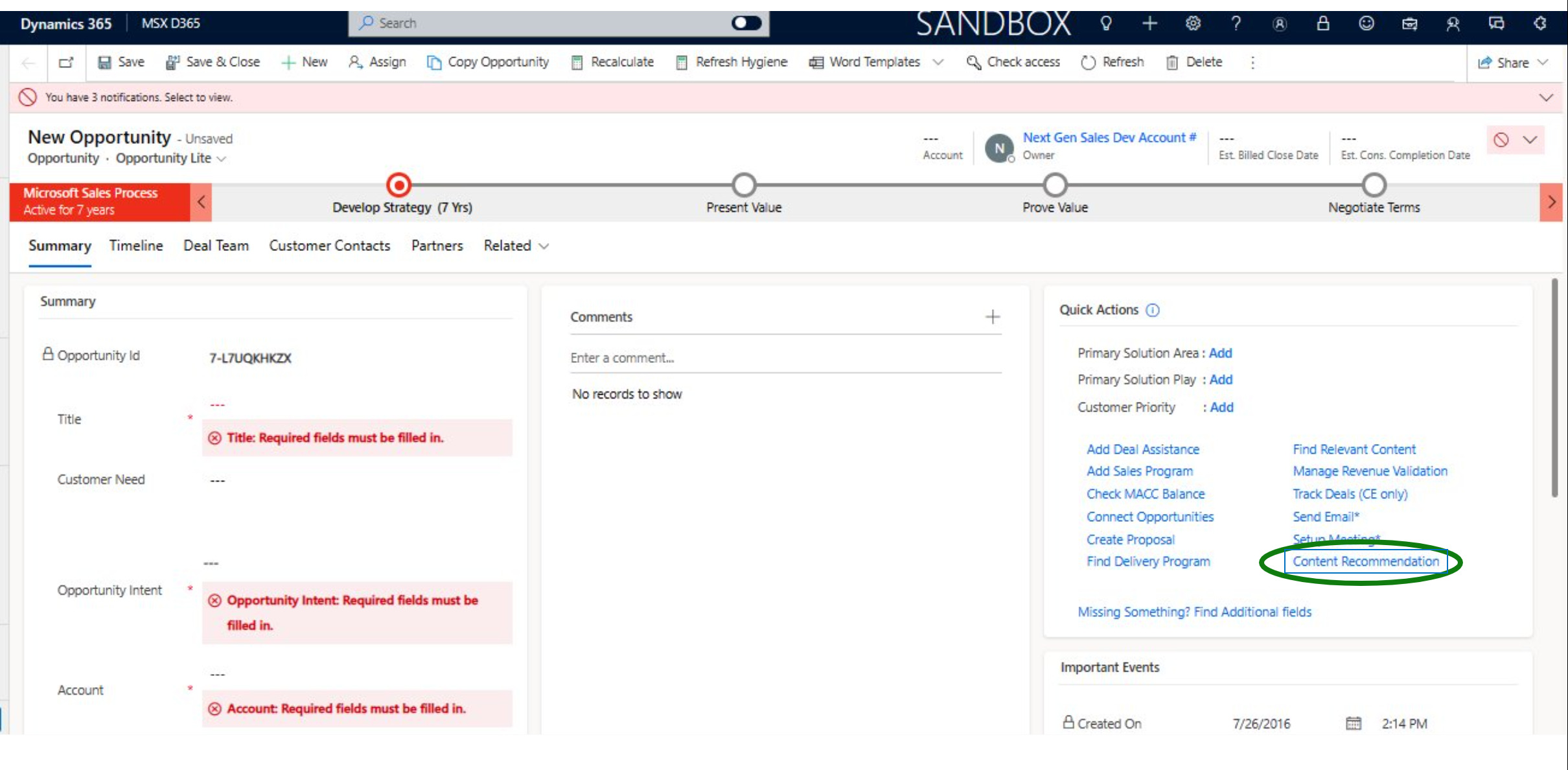}}
\vspace{0.5cm}
\centerline{\includegraphics[width=2\columnwidth]{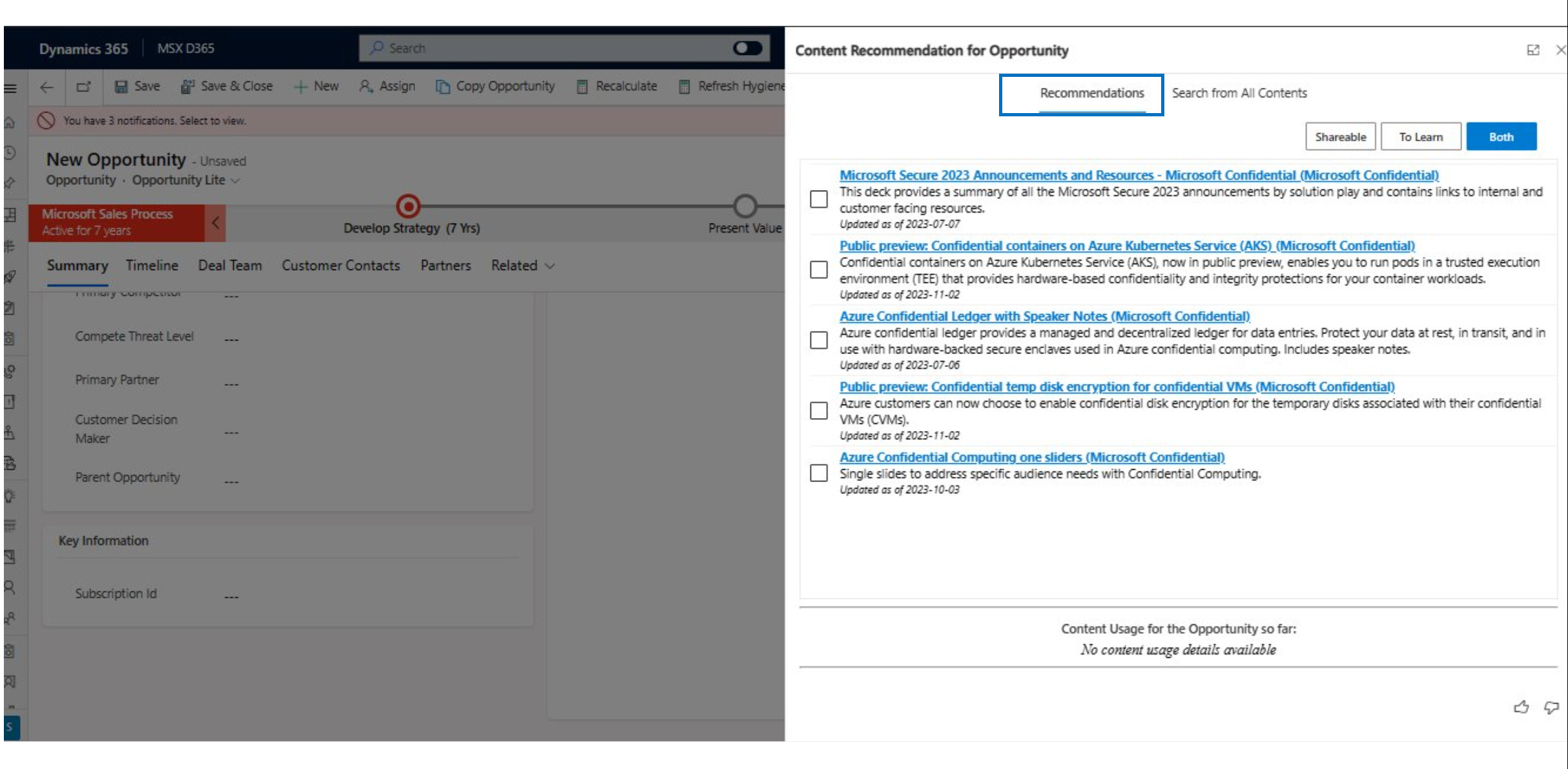}}
\caption{Integration into the Microsoft Seller Experience (MSX) UI. The \textbf{top} figure shows a general view of the interface sellers are interacting with to manage opportunities.  Notice in the bottom right a link highlighted by a dark-green ellipse.  By clicking on this link, sellers can view and browse the recommended Seismic content specific to this opportunity as illustrated in the \textbf{bottom} figure and the highlighted blue rectangle.  Immediately to the right, one can see the ``search for all contents'' tab which invokes the copilot model developed in~\cite{copilotMSX} for a more general interactive question-answering experience.}
\label{fig:MSXUI}
\end{center}
\end{figure*}

We show in~Fig.~\ref{fig:MSXUI} an illustration of how the recommender system is integrated into the MSX UI.

\end{document}